\documentclass[12pt]{article}

\usepackage[active]{srcltx}
\usepackage[a4paper]{geometry}
\usepackage{amsthm,amsfonts,amsmath,amssymb}
\usepackage{array}
\usepackage{booktabs}
\usepackage{bbm}
\usepackage{changes}
\usepackage{cancel}
\usepackage{dsfont}
\usepackage{esint}
\usepackage{graphicx}
\usepackage{hyperref} \hypersetup{colorlinks=true,pdfborder={0 0 0}}
\usepackage{paralist}
\usepackage{tikz}
\usepackage[utf8]{inputenc}
\usepackage{ulem}
\usepackage{verbatim}

\def \be{\begin{equation}}
\def \ee{\end{equation}}

\newcommand{\der}[2]{\frac{d #1}{d #2 }}
\newcommand{\scp}[2]{\mbf{ #1}\cdot\mbf{ #2 }}
\newcommand{\e}{\varepsilon}

\newcommand{\pder}[2]{\frac{\partial #1}{\partial #2 }}

\newcommand{\mbf}[1]{\mathbf{ #1} }
\newcommand{\average}[1]{\mathbb{E}\left( #1\right)}

\DeclareMathOperator{\sgn}{sgn}
\DeclareMathAlphabet\mathbfcal{OMS}{cmsy}{b}{n}

\title{A relativistically exact Eikonal equation for optical fibers with application to   adiabatically deforming
 ring interferometers}

\begin{document}
\author{Joseph Avron and Oded Kenneth \\
 Deptartment of Physics, Technion, Israel}

\maketitle

\begin{abstract}
We derive the relativistically exact Eikonal equation for  ring interferometers undergoing   deformations. For ring interferometers that undergo slow deformation we describe the two leading terms in the adiabatic expansion of the phase shift.  The leading term 
is independent of the refraction index  $n$  and is given by a line  integral  generalizing results going back to Sagnac 
\cite{sagnac1913,wang2004,Ori}  for non-deforming interferometers to all orders in {$\beta=|\mbf{v}|/c$}.
In the non-relativistic limit {this term} is $O(\beta)$. 
The next term in the adiabaticity has the form of a double integral, it is of 
order $\beta^0$ and depends on the refractive index $n$. It accounts for non-reciprocity due to changing circumstances in the fiber.  The adiabatic correction is often comparable to the Sagnac term.  In particular, this is the case in Fizeau's interferometer.
Besides providing a mathematical framework that puts all ring interferometers  under a single umbrella, our results generalize and strengthen results of \cite{Ori, shupe} to fibers with chromatic dispersion. 
\end{abstract}
\section{Introduction and summary of results}

A ring interferometer has a light source with prescribed frequency which produces two counter propagating waves which interfere  after completing a full cycle.  
An example is Sagnac interferometer \cite{sagnac1913} originally devised to measure the velocity of light relative to the ether but today is perhaps best 
known because of its close relative, the Ring Laser Gyro \cite{rlg}, which literally impacted everyday life.  Modern fiber-optics ring interferometers have multiple uses as gyroscopes \cite{Vali}, strain sensors \cite{strain} temperatures sensor \cite{temperature}, filters etc. They have been a fertile ground for addressing basic physics 
and technological issues,   \cite{ajp,arditty,hasselbach,matter-wave,gravity,yale,Ori,schiller,wang2004} and have been reviewed  in {e.g.}
\cite{ correct,post1967,tests}.

  The study of the wave equation in moving dielectrics is fraught with both conceptual and technical difficulties \cite{menegozzi,shiozawa}. A simplification occurs in the
the short wavelength, high frequency limit, which is described by the Eikonal equation 
\cite{arnold}.   The main drawback of the Eikonal is that it disregards  backscattering. 
This is sometimes important \cite{limitation}, but more often the Eikonal 
gives an adequate description of the interference in ring interferometers.  
As far as relativity is concerned, the Eikonal equation is, in principle, exact.    

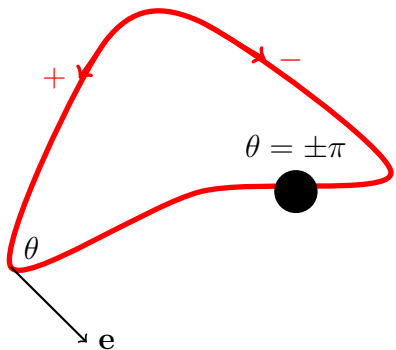
\begin{figure}[h!]
\centering{} \begin{tikzpicture}[scale=2.5,line width=2pt,red]
\pgfplothandlerclosedcurve
\pgfplotstreamstart
\pgfplotstreampoint{\pgfpoint{0cm}{0cm}}
\pgfplotstreampoint{\pgfpoint{.5 cm}{1.2cm}}
\pgfplotstreampoint{\pgfpoint{1.cm}{1.3cm}}
\pgfplotstreampoint{\pgfpoint{2cm}{0.5cm}}
\pgfplotstreampoint{\pgfpoint{1cm}{0.4cm}}
\pgfplotstreamend
\pgfusepath{stroke}
\node [right,black] at (0,.1) {$\theta$};
\node [above,black] at (1.5,.5) {$\theta=\pm\pi$};
\draw[black,thick,->] (0,0)-- (.4,-.4) node [right] {$\mathbf e$};
\draw[fill,black] (1.5,.4) circle (.1);
\draw [->,red,ultra thick] (.5,1.2)--(.36,1.) node[left] {$+$};
\draw [->,red,ultra thick] (1,1.3)--(1.34,1.1) node[right] {$-$};
\end{tikzpicture} 
\caption{The ring interferometer is represents by  parametrized curve ${\mathbf{x}}(\theta,t)$ which is  a closed loop in space. $\theta$ designates a fixed 
material points on the
fiber.
 $\mathbf{e}(\theta,t)$ is the tangent 
at a point $\theta$. The two counter-propagating beams
are marked with the red arrows and the $\pm$ signs. The black dot
represents the co-moving  light source and detector {both} located at $\theta=\pm\pi$. }
\label{f:3} 
\end{figure}

We construct the Eikonal equation for deformable fiber interferometers of arbitrary shape,  Fig.~\ref{f:3},  moving at velocities that may be relativistic 
($\beta=|\mbf{v}|/c\approx 1$), while bending and stretching.  
We then proceed to  describe an expansion of the phase shift {for adiabatically deforming fibers}.

To describe our results we need to introduce some notation.
A thin ring fiber is  naturally modeled by a one dimensional closed curve  
$\bold x (t,\theta)\in\mathbb{R}^3$ where  the points $\theta=\pm\pi$ are identified.  
{ It is convenient to choose 
parametrization where $\theta$ labels the material points of the fiber\footnote{{For general parametrization see Appendix \ref{a:stationary}}}. The co-moving  light source and detector are located at $\theta=\pm\pi$. }
$t$ is the Lab time coordinates.

The differential
\be\label{e:dx}
d\mbf{x}= \mbf{v}\, dt + \mbf{e}\, d\theta,
\ee
gives $\mbf{v}(t,\theta)$, the velocity of the point $\theta$ in the lab,  and  $\mbf{e}(t,\theta)$ the tangent to the curve. The length of the segment $d\theta$ in the Lab is
\be
d\ell=|\mbf{e}|\, d\theta\,.
\ee

The Eikonal is a non-linear first order PDE that governs the evolution of  the phase of the wave $\phi(t,\theta)$: 
\be\label{e:eik1}
\pm
 \pder{\phi}{\theta}= K_\pm(t,\theta,\omega), \quad  {\omega}= -\pder{\phi}{t}
\ee
The $\pm$ sign distinguishes  the two counter-propagating waves:
The $(+)$ wave propagate with $\theta$ increasing from $-\pi$ to $\pi$ and the $(-)$ wave propagates with $\theta$ decreasing from $\pi$ to $-\pi$.  
The explicit form of $K_\pm$, given in Eq.~(\ref{e:H}) below,  need not concern us at this point. 

The phases $d\phi_\pm$, accumulated by the counter-propagating waves as they traverse the interval $d\theta$ are, in general, different. The phase difference $\delta\phi=d\phi_+- d\phi_-\neq 0$ is known as ``non-reciprocity''.  It has two origins: 
{First, $K_+\neq K_-$ due to the dependence on the wave propagation direction
and moreover, since the two waves visit the interval $d\theta$ at different times 
$t_+(\theta)\neq t_-(\theta)$, the value of physical parameters making up $K_\pm(t,\theta)$ may have changed.}

 The amplitude of the wave is governed by suitable transport equations \cite{Balyan} and is slowly varying   when the frequency is high.  We shall not study it here.  

Assuming that the amplitudes of the $(\pm)$ waves  are the same, the detector output at the  time of detection $t$ is proportional to  
\be
|e^{i\phi_+(t,\pi)}+e^{i\phi_-(t,-\pi)}|^2=2(1+ \cos\big(\Delta\phi (t)\big), \quad  \Delta\phi(t)=\phi_+(t,\pi)-\phi_-(t,-\pi)
\ee
Our goal is to derive an expansion for the phase shift $\Delta\phi(t)$ for fibers that deform adiabatically


Before describing the adiabatic expansion we need to describe  the notion of adiabaticity. 
The $\pm$ waves visit the interval $d\theta$ at different times, $t_\pm(\theta)$. Consequently, the phase difference $\delta\phi=d\phi_+- d\phi_-$ may depend on the time lapse $\delta t =t_+(\theta)-t_{-}(\theta)$. For example, the interval $d\theta$ may stretch so that the length $d\ell=|\mbf{e}|\, d\theta$ seen by the two waves is different. Similarly 
{the refraction index} $n(\theta,t,\omega)$ or $\beta$ may change between the two visits.
We say that the motion of the fiber is adiabatic if all such changes, for all values $\theta$,   are small. This is the case if $K_\pm$ changes little  in the time $\tau$ it takes light to complete a cycle.  
A natural dimensionless measure of adiabticity  is  then
{
\be
 \e=\tau\partial_t\log(K_\pm)
\ee
{If $n$ is time independent and $\beta\ll 1$ then $\tau\partial_t \log(K_\pm)\approx n\sigma|\mbf{e}|$ where $\sigma$ is the rate of stretching of the fiber, see Eq.~(\ref{e:sig2}).}
The adiabatic regime is $|\e|\ll 1$.

For large interferometers, and for interferometers with a large number of  coils \cite{lefevre}, $\tau$ need not be small compared with the  time scale {of the deformation and} the assumption of adiabaticity may fail.

Our main result is an expansion of the phase shift in powers of $\e$
\be
\Delta\phi=(\Delta\phi)^{(0)}+(\Delta\phi)^{(1)}+\dots
\ee
where $(\Delta\phi)^{(j)}=O(\e^j)$. The leading term, $(\Delta\phi)^{(0)}(t)$, is given by a line integral at the time of detection $t$:
\begin{align}\label{e:stat}
(\Delta\phi)^{(0)}
&= -\frac{2\omega_0}{c^2} \int_{-\pi}^\pi \,\gamma^2(\theta) \,\mbf{v}(\theta) \cdot  \mbf{e}(\theta)\,  {d\theta}\nonumber\\
&= -\frac{2\omega_0}{c^2} \int_{0}^L \,\gamma^2(\ell) \,\mbf{v}(\ell) \cdot
     {d\boldsymbol{\ell}}
	\end{align}
where, for the sake of typographical simplicity, we have omitted the common argument $t$ everywhere. The {length} variable $\ell$ {is related to $\theta$} by 
$\ell=\int_{-\pi}^\theta |\mbf{e}(\theta')|\,d\theta'$.  
 $\gamma=(1-\beta^2)^{-1/2}$ and $L$ is the {total} length of the fiber (at time $t$). 
Remarkably, $ (\Delta\phi)^{(0)} $ is independent of the (constitutive) {dispersion} 
relation $n$ {\em  to all orders in $\beta$}.
$\omega_0$ is  the  frequency of the source measured by the lab clocks.
 
In the approximation $\gamma\approx 1$, which is always  the case in practice, one recovers  the result of \cite{Ori}. 
If, in addition, the interferometers {moves as a rigid body}, the application of Stokes formula  recovers the standard Sagnac {area law} \cite{wang2004,post1967,correct}. 

 The first order correction in adiabaticity,  $(\Delta\phi)^{(1)}(t)$, has the form of a double integral along the fiber evaluated at the time of detection $t$.  In the non-relativistic limit we  find for $(\Delta\phi)^{(1)}(t)$:
\begin{align}\label{e:2nd}
(\Delta\phi)^{(1)}
=\frac \omega {c^2}\int_{-\pi}^\pi\int_{-\pi}^\pi \, |\mbf{e}(\theta)|d\theta\, |\mbf{e}(\theta')|d\theta'\,  \partial_\omega\big(\omega n(\theta')\big)\,
\big(\partial_{t} n(\theta)+n(\theta)\sigma(\theta)\big)\,
\sgn(\theta'-\theta)
\end{align}
We suppressed the common argument $t$ on both sides. 
$\sigma=\partial_t\log|\mbf{e}(\theta,t)|$ is the (non-relativistic) stretch rate of the fiber, see Eq.~(\ref{e:sig2}) below.   
Note that while Eq.~(\ref{e:stat}) is independent of $n$, Eq.~(\ref{e:2nd})   depends quadratically on $n$.   

 
$(\Delta\phi)^{(0)}=O(\beta\e^0)$ while $(\Delta\phi)^{(1)}=O(\e\beta^0)$.
In principle,  $\e$ and $\beta$ are independent parameters
\footnote{If the length scale of the deformations is comparable to the length of the fiber
and the velocities associated with them are comparable to the rigid body part of the velocity than $\e\sim\beta$.}
When $\varepsilon\ll \beta$,  $(\Delta\phi)^{(1)}$  is a  small correction to  
$(\Delta\phi)^{(0)}$ and in the opposite case  $\varepsilon\gg \beta$,  $(\Delta\phi)^{(0)}$  is a  small correction to  $(\Delta\phi)^{(1)}$.  
When $\e\sim\beta$ the two terms are comparable. This is the case in Fizeau's interferometer \cite{fizeau}, shown schematically in Fig. \ref{f:fizeau}. Although this is 
not a fiber interferometer, the theory still applies and as we shall show in section \ref{s:fizeau} 
von Laue's classical  formula \cite{laue} for the phase shift in Fizeau is  simply the sum of the two terms
\begin{equation}
\frac{2\omega}{c^{2}} (V L)\,\big(n\partial_\omega(\omega n)-1\big)=(\Delta\phi)^{(0)}+(\Delta\phi)^{(1)}
\label{e:f}
\end{equation}where $V$ is the velocity of the flow and $L$ the length of the pipe.  
 \cite{arditty,Ori}. 

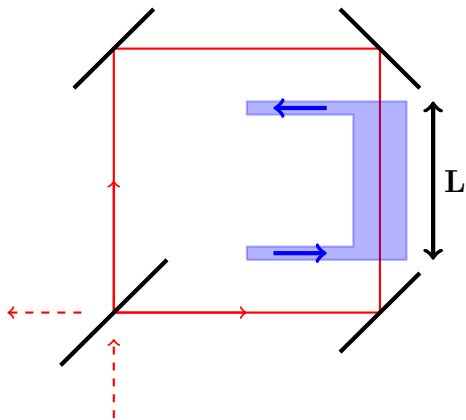
\begin{figure}[h!]
\begin{centering}
\begin{tikzpicture}[scale=3.5]
\draw[ thick,red] (0,0)-- (1,0)--(1,1)--(0,1)--(0,0);
\draw[ thick,red,->] (0,0)-- (.5,0);
\draw[ thick,red,->] (0,0)-- (0,.5);
\draw[ultra thick]  (1-.15,0-.15)--(1+.15,0+.15);
\draw[ultra thick]  (0-.15,1-.15)--(+.15,1+.15);
\draw[ultra thick]  (1+.15,1-.15)--(1-.15,1+.15);
\draw[ultra thick,black] (-.2,-.2)-- (.2,.2);
\draw[ thick,red,->,dashed] (0,-.4)-- (0,-.1);
\draw[ thick,red,<-,dashed] (-.4,0)-- (-.1,0);
\draw[thick,fill, blue,opacity=.3](.5,.2)--(1.1,.2)-- (1.1,.8)--(.5,0.8)--(.5,0.8-.05)--
(.9,0.8-.05)--(.9,0.2+.05)--(.5,0.2+.05)--(.5,.2);
\draw[ultra thick,blue,->] (.6,0.225)-- (.8,.225);
\draw[ultra thick,blue,<-] (.6,0.225+.55)-- (.8,.225+.55);
\draw[ultra thick,black,<->] (1.+.2,0.2)-- (1.+.2,.8);
\node [right] at (1.2,.5) {$\bold L$};
\end{tikzpicture}
\par\end{centering}

\centering{}\protect\caption{Schematic Fizeau interferometer. 
The three mirrors and beams splitter are at rest in the lab. The two counter-propagating beams are denoted by the red arrows. 
Fluid, say water, is flowing in one arm of the interferometer so one beam is moving with the flow and the other against it.}
\label{f:fizeau} 
\end{figure}

\section{Space-time geometry of a moving ring }\label{s:setup}

A moving curve in Minkowski space-time can be described by a vector valued function of two variables: $\bold x (t,\theta)\in\mathbb{R}^3$.  There is freedom in choosing  parameterization for the curve. A convenient parameterization is to choose $t$ to be the Lab time coordinate and $\theta$ labeling material points of the fiber.  The velocity and tangent to the curve are given in Eq.(\ref{e:dx}). It follows that 
\be
\partial_t\mbf{e}=\partial_\theta\mbf{v}
\ee
The (lab frame) line element is $d\boldsymbol{\ell}=\mbf{e}\, d\theta$. 
Since $\mbf{v}$ is a velocity of a material point on the fiber  $\scp{v}{v}<1$ in units
where $c=1$ {which we henceforth use}.

If $\mbf{e}$ is parallel to $\mbf{v}$, the proper length\footnote{As measured by an observer co-moving with the material point of the fiber labeled by $\theta$.}  of the segment $d\ell'$ is related to the length $d\ell$ by Lorentz contraction, 
$ \gamma d\ell=d\ell'$. If, however, $\mbf{e\cdot v}=0$ there is no contraction and $d\ell=d\ell'$.  In general, $d\ell$ and $d\ell'$ are related by
 \be\label{e:ell'}
 \frac{d\ell}{\gamma_\perp} =
\frac {d\ell'}{\gamma}, 
 \quad
   \gamma=\frac 1 {\sqrt{1-\mbf{v}^2}},\quad \gamma_\perp=\frac 1 {\sqrt{1-(\mbf{v\times\hat e})^2}}
   \ee
This can be seen from
\begin{align}\label{e:ell'}
(d\ell')^2&=\left(d\ell'_\perp\right)^2+\left(d\ell'_\|\right)^2\nonumber \\
&=
\left(d\ell_\perp\right)^2+{\left(d\ell_\|\right)^2\over 1-\mbf{v}^2}\nonumber \\
&=\gamma^2\left((d\ell)^2-\mbf{v}^2 (d\ell_\perp)^2\right)\nonumber \\
&=\gamma^2(d\ell)^2(1-\mbf{v}^2_\perp)=
\left({\gamma  d\ell\over \gamma_\perp}\right)^2
\end{align}

The  fibers are allowed to stretch. The stretch rate $\sigma$ is naturally defined as the rate of change  of proper length, measured in  proper time (as measured by an observer co-movong with the point $\theta$ of the fiber):
\be\label{e:sigma}
\sigma=\gamma \partial_t\log \der{\ell'}{\theta}, \quad \der{\ell'}{\theta}= \frac \gamma{\gamma_\perp}|\mathbf{e}|
\ee
To leading order in $\beta$
\be\label{e:sig2}
\sigma\approx \partial_t\log |\mbf{e}|=\frac{\mbf{e}\cdot\mbf{\dot e}}{\mbf{e}\cdot\mbf{ e}}=\frac{\mbf{e}\cdot\partial_\theta\mbf{ v}}{\mbf{e}\cdot\mbf{ e}}
\ee
No-stretching means $\sigma=0$. 

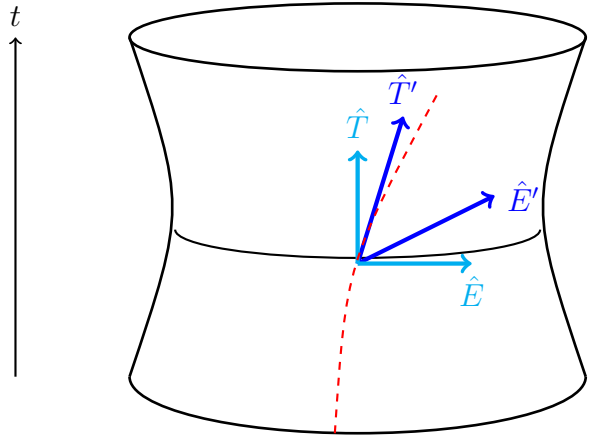
\begin{figure}[h!]
\centering{} 
\begin{tikzpicture}[scale=1.5,line width=1pt,black]
\begin{scope}
 \draw (2,0) arc (0:-180:2cm and .5cm);
  \draw (0,3) ellipse (2cm and .3cm);
  \draw[thick] (1.6,1.3) arc (0:-180:1.6cm and .25cm);
  \draw (-2,0) .. controls (-1.5,1.5) .. (-2,3);
  \draw (2,0) .. controls (1.5,1.5) .. (2,3);
   \draw [blue,->,ultra thick](0,1.) -- (.4,2.3) node [above] {$\hat T'$};
    \draw [blue,->,ultra thick](0,1.) -- (1.2,1.6) node [right] {$\hat E'$};
        \draw [cyan,->,ultra thick](0,1.) -- (1,1.) node [below] {$\hat E$};
            \draw [cyan,->,ultra thick](0,1.) -- (0,2.) node [above] {$\hat T$};
     \draw [red,thick,dashed] (-.2,-.5) .. controls (-.1,1.) .. (.7,2.5);
     \draw [thick, ->] (-3,0)--(-3,3) node [above] {$t$};
     \end{scope}
\end{tikzpicture} 
\caption{The figure represents the two dimensional world-tube $\cal M$ associated with a moving closed curve in Minkowski  space-time.  
The red curve is the world line of a fixed material  point  of the fiber labeled by $\theta$.  The Lorentz orthonormal frame $(\hat T, \hat E)$ distinguishes the time and space axis in the lab. Similarly,  $(\hat T',\hat E')$ distinguishes the time and space axis in the co-moving frame attached to $\theta$. }
\label{f:m} 
\end{figure}
  The world line of a fixed  material point {labeled by $\theta$} is 
\be
X(t,\theta)= (t, \bold x (t,\theta)), \quad -\infty<t<\infty\,,
\ee
The union of the world lines for all $\theta$ gives a two dimensional tube {shown in Fig. \ref{f:m}}
\be
{\cal M}=\big\{X=(t,\mbf{x}(t,\theta))\big| \quad \theta\in [0,2\pi], \quad t\in\mathbb{R}\big\}
\ee
The tangent vectors to $\cal M$ are spanned  by 
$T'=\partial_tX=(1,\mbf{v})$ and $E=\partial_\theta X=(0,\mbf{e})$. 
The scalar product is induced from the  
embedding 4-D Minkowski space:
\begin{align}
T'\cdot T'&=\partial_tX\cdot \partial_tX= 1-\scp{v}{v}
\nonumber \\
E \cdot E&= \partial_\theta X\cdot \partial_\theta X=-\scp{  e}{ e} \\
T'\cdot E&= \partial_\theta X\cdot \partial_t           X=-\scp{ e}{v}\nonumber
\end{align}
The metric on ${\cal M}$, in the coordinates $(\theta,t)$, is given by
\be
dX\cdot dX=(1-\mbf{v}^2)(dt)^2-2\,\scp{e}{v}\, dt\, d\theta-\mbf{e}^2\, (d\theta)^2
\ee 
Note that the coordinates are not Lorentz orthogonal.

$E=\partial_\theta X$ is the direction of simultaneity in the lab.
We shall denote by $\hat E$ the  normalized version of $E$.
The time  direction in a co-moving frame is given by $T'=\partial_tX$.  
Its normalized version  $\widehat{T'} =\gamma T'$ corresponds to the 4-velocity of the material point with fixed $\theta$. Explicitly
\be\label{e:T'}
 \widehat{T'}=\gamma \partial_t X=\gamma(1,\mbf{v})\,,\quad \widehat{E}=\frac { \partial_\theta X}{|\mbf{e}|}=(0,\mbf{\hat e}),
\ee 
The spatial direction of the co-moving Lorentz frame is the tangent vector $E'$ which is Minkowski orthogonal to $T'$:   
 \be\label{e:E'}
 E'=  E-(E\cdot \widehat T') ~ \widehat T' =\partial_\theta X+{ \gamma^2}  \scp{v}{  e}  \, \partial_t X
 \ee
  The Minkowski length of $E'$ is
 \begin{align}\label{e:lamb}
 E'\cdot E'
 &= E\cdot E - (E\cdot \widehat T')^2\nonumber\\
 &=-\mbf {{e}}^2- {\gamma^2}( \mbf{{e}} \cdot \mbf{v})^2\nonumber
 \\
 &=- \gamma^2\left(\mbf{e}^2-(\mbf{v\times {e}})^2\right)\nonumber\\
 &=-\frac{\gamma^2}{\gamma_\perp^2}\mbf{e}^2
 \nonumber\\
 &=-
 \left(\der{\ell'}{\theta}\right)^2
 \end{align}
For the sake of completeness we note that the tangent vector to $\cal M$ associated with the Lab time is
 \begin{align}
T&= T'- (\widehat E\cdot T') {\widehat E}=(1,0)
 \end{align}
  $(\widehat T,\widehat E)$ and $(\widehat T',\widehat E')$ are Lorentz orthogonal frames related by standard Lorentz transformation.  In contrast,  the frame
$(\widehat T',\widehat E)$ associated with the coordinates  $(t,\theta)$  is not Lorentz orthogonal and so is not related to 
$(\widehat T',\widehat E')$ by the standard Lorentz transformation.


 \section{The Eikonal}\label{s:eikonal}
{   
 In a local co-moving frame the wave vector $k'$ and frequency $\omega'$ are related by the dispersion relation\footnote{Allowing $t$ dependence of $n$ implicitly also allows it to depend on the curvature of the fiber.}:
\be\label{e:11}
{ k' =\pm n(t,\theta,\omega') \omega'}
\ee 
These local inertial frames  do not provide a global frame on $\cal M$ because clocks associated with different local frames at different $\theta$ 
can not be synchronized.  We need to translate  Eq.~(\ref{e:11}) to a relation between 
$\partial_t\phi$ and $\partial_\theta\phi$. 

By definition, 
\be\label{e:omega'}
\omega'= -\nabla_{\widehat T'}\phi,\quad 
  k'=\nabla_{\widehat E'}\phi
\ee
Using Eq.~(\ref{e:T'}) we thus have
\be\label{e:omega'1} \omega'=-\gamma\partial_t\phi=\gamma\omega  \ee
and using Eqs.~(\ref{e:E'},\ref{e:lamb})
\be\label{e:k'1}  k'={\der{\theta}{\ell'}}(\partial_\theta+\gamma^2\scp{v}{  e}\partial_t)\phi \ee
{\bf Remark:} As the time derivative $\partial_t=(\partial_t)_\theta$ is taken at constant value of the comoving coordinate $\theta$, it has the physical meaning of a derivative taken 
with the flow. 
Had we used the lab coordinate $(x,t)$ it would have taken the Eulerian form
$\partial_t+\scp{v}{\nabla}$ and Eq.~(\ref{e:omega'1}) would have taken the standard form 
of a Doppler shift. The comoving coordinates $(\theta,t)$ are however not inertial
and hence the transformation takes a different form.

Substituting $\omega'$ and $k'$ from Eqs.~(\ref{e:omega'1},\ref{e:k'1}) 
in Eq.~(\ref{e:11}) we find:
\be
 \der{\theta}{\ell'}\left(\partial_{\theta}+{ \gamma ^2}  \scp{v}{{e}}  \, \partial_t\right)\phi=\mp \gamma \, n\,\partial_t\phi 
,\quad    n=n\left(t,\theta,\gamma \omega\right) 
\ee
}

Rearranging gives the Eikonal equation:
\be\label{e:eik}
\pm
 \pder{\phi}{\theta}= K_\pm(t,\theta,\omega), \quad  {\omega}= -\pder{\phi}{t}
\ee
where 
\begin{align}\label{e:H}
 K_\pm(t,\theta,\omega,)&=  K_1(t,\theta,\omega,)\pm  K_2(t,\theta,\omega,)\nonumber \\
 &=\gamma^2 \frac{\omega}c \left( \, \frac{ n(t,\theta,\gamma\omega)} {\gamma_\perp} \pm  
\frac{\scp{v}{ \hat e}}c\right){|\mbf{e}|} 
\end{align}
 We have now reintroduced $c$.  $K_\pm$  is a dimensionless version of the wave number encoding all the relevant information on the fiber and its motion.
 
In the non-relativistic limit where  $\gamma,\gamma_\perp\approx 1$
\be\label{e:H1}
K_\pm\approx \frac  \omega c\left( n \pm \frac{ \scp{v}{\hat e}} {c}\right){|\mbf{e}|} 
\ee

The Eikonal, {Eq.~(\ref{e:eik})}, is a non-linear, first order  PDE. In the case that {$n$} is non-dispersive, it simplifies to a linear PDE.   In either case, the PDE can be solved by the 
method of characteristics, which reduces the problem of solving a PDE to solving a set of coupled ODE's \cite{john}.
{We describe this reduction in the next subsection.}
\subsection{The Hamiltonian system}\label{s:H}
 The Eikonal equation Eq.~(\ref{e:eik}) has the form of Hamilton Jacobi equation of mechanics.  Table \ref{table} gives the dictionary that translates  wave properties to mechanical properties.  This allows to easily write the characteristic equations and reduces solving the  PDE for the phase $\phi_\pm(\theta,t)$ to a problem in mechanics.

\begin{table}[h!]
\centering
\begin{tabular}{| l c| l r| }
\hline
  Mechanics && Eikonal& \\
  \hline\hline
  Phase space & (x,p)&Time-Frequency plane &$(t,\omega)$ \\
  \hline
   Position& $x$& Time  & $t$\\
  \hline
  Momentum& $p$&Frequency& $\omega$\\
  \hline
  Time& $t$& Fiber coordinate & $\pm\theta$\\
  \hline
   Action& $S(x,t)$ &  {(minus)} Phase & $-\phi(t,\theta)$ \\
  \hline
  Hamiltonian& $H(x,p,t)$ & Dimensionless wave number& $K_\pm(t,\theta,\omega)$  \\
  \hline
  Lagrangian &L(t,x,v)&$\cal L$ &$\omega \partial_\omega K-K$\\
  \hline
 \end{tabular}
 \caption{The correspondence between Hamilton-Jacobi equations in mechanics and the Eikonal equation. }\label{table}
 \end{table}
 
The phase $\phi_\pm(t,\theta)$ is the analog of the action $S(x,t)$ which can be determined by solving the Hamilton equations for $(x,p)$ and then integrating 
\be
dS=p dx -H dt= (p\dot x -H) dt= Ldt
\ee
 along the classical path.  We shall do precisely the same thing for $\phi_\pm$. 

The analog of phase space in the context of the Eikonal is the time-frequency plane $(t,\omega)$ and the analog of time in mechanics is the coordinate 
$\theta$ of the fiber. Hence, the analog of Hamilton equations are:
\begin{align}\label{e:ham}
\pm\der{t_\pm}{\theta}=\partial_{\omega} K_\pm \, , \quad 
\pm\der{\omega_\pm}{\theta}=- \partial_tK_\pm 
\end{align}
The equation for $\omega_\pm$ may be interpreted as Doppler shifts along the fiber

\begin{figure}[h!]
\begin{centering}
\begin{tikzpicture}[scale=3]
\draw[dashed] (0,0)--(1.2,0);
\draw[thick,->]  node [below] at (0,-1.3) {$\theta=-\pi$}  (0,-1.3)--(0,.6) node [above] {$t$};
\draw[dashed]   (1.2,-1.3)--(1.2,.5) ;
\node [below] at (1.2,-1.3)  {$\theta=\pi$};
\draw[dashed,<-] (0,-1.)--(1.2,-1.) ;
\draw[dashed,<-] (0,-1.2)--(1.2,-1.2) ;
\draw[ultra thick,->,red]  (0,-1)--(1.2,0);
\draw[ultra thick,->,blue]  (1.2,-1.2)--(0,0);
\node [below]  at (1,-.25) {$t_+$};
\node [below]  at (.25,-.35) {$t_-$};
\node [above left]  at (0,0) {$t_-(-\pi)=t_d$};
\node [left]  at (0,-1) {$t_+(-\pi)$};
\node [right]  at (1.2,-1.2) {$t_-(\pi)$};
\node [above right] at (1.2,0) {$t_+(\pi)=t_d$};
\draw[thick,->] (0,-.7)--(1.5,-.7) node [right] {$\theta$};
\end{tikzpicture} 
\caption{The  characteristics for the $+$ wave is the red arrow and for the $-$ wave the blue arrow. Both terminate at the detector simultaneously at Lab time $t_d$. 
The characteristics are parametrized  by $\theta$. The interfering waves have different times of emission. 
This difference is one source for the phase shift. 
The other source is the evolution of the phase  along the  characteristics.  }
\label{f:schedule} 
\end{centering}
\end{figure}
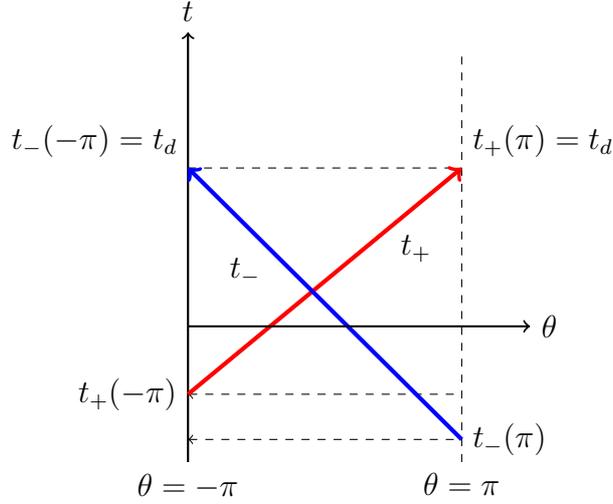

The evolution of the phase $\phi$ along the fiber is now
\be\label{e:phi}
d\phi_\pm=\mp{\cal L}_\pm \, d\theta
\ee
where ${\cal L}_\pm$ is the Legendre transform of $K_\pm$:
\begin{align}\label{t46r5}
{\cal L}_\pm &=\pm\omega \der{t_\pm}{\theta}-K_\pm\nonumber \\
&=\omega \partial_\omega K_\pm-K_\pm\nonumber \\
&=\omega \partial_\omega K_1-K_1\nonumber \\
&=\frac{\omega^2} c\gamma\,\big(\partial_\omega n\big)\, \der{\ell^\prime}{\theta}
\end{align}
In the third identity we used {the fact that $K_2$ is linear in $\omega$ so its Legendre transform vanishes identically.}
This shows that ${\cal L}_+={\cal L}_-={\cal L}$. In the absence of dispersion\footnote{This reflects the fact that in the absence of dispersion, {$t_\pm(\theta)$ equals} the time schedule of a constant phase along the fiber. }.
  ${\cal L}=0$.
To first order in $\beta$ 
\be
{\cal L} \approx  \frac{\omega^2} c
\big(\partial_\omega n\big)
\, |\mbf{e}|
\ee
Once we find the solutions $(t_\pm(\theta),\omega_\pm(\theta))$ for the Hamiltonian system, the phase $\phi_\pm$ can be obtained by integrating  
Eq.~(\ref{e:phi}) along the trajectory.

\subsection{The boundary conditions}\label{s:bc}
We now turn to the boundary conditions for the Hamiltonian system,  Eqs.~(\ref{e:ham}), governing the flow of the {phase space}  points $(t_\pm,\omega_\pm)$   
 and $\phi_\pm$, Eqs.~(\ref{e:phi},\ref{t46r5}). 

 Consider  Fig. \ref{f:schedule}. We are interested in solutions $t_\pm(\theta)$  that terminate simultaneously at the detector at $t_d$.     
This  imposes  {\em final} conditions  on $t_\pm(\theta)$:
 \be\label{e:tbc}
 t_+(\pi)=t_-(-\pi)= t_d 
\ee
Note that the emission times are not specified and the  $\pm$  waves may have  different emission times:
\be
{t_{e+}=t_+(-\pi),\quad t_{e-}=t_-(\pi)}
\ee
 
The second boundary condition fixes  $\omega_\pm(\theta)$ to be the frequency of the source {$\omega_0(t_e)$} at the time of emission:
{
\be\label{e:obc}
\omega_\pm(\mp\pi)={\omega_0(t_{e\pm})}=\left.\frac{\omega_0'}{\gamma_\pi}\right|_{t=t_{e\pm}},\quad  \gamma_\pi(t)=\gamma(t,\pi)
\ee
Where we assumed that the source has a constant frequency $\omega_0'$ in its own rest frame.
The phase of the source is 
\be\label{e:g}
\Phi=-\omega_0' t'=-\omega_0' \int^t\frac{ds}{\gamma_\pi(s)}
\ee
and one may therefore write the boundary condition on $\omega_\pm$  as
\be\label{e:mbcw}
\omega_\pm(\mp\pi)=
{-\left(\der{\Phi}{t}\right)\big(t_{e\pm}\big)}
\ee
}

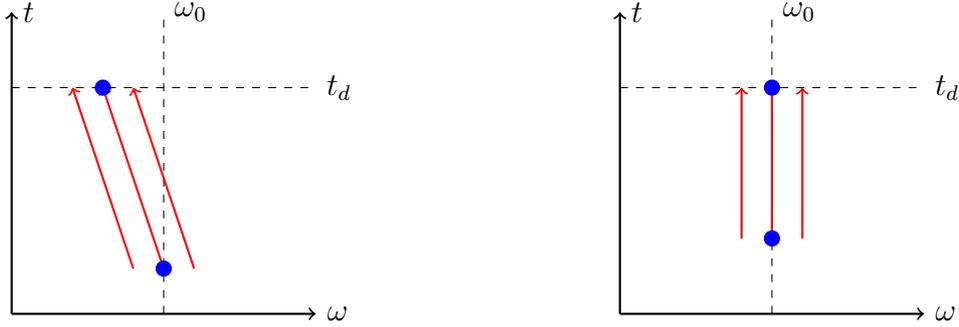
\begin{figure}[h!]
\centering{} \begin{tikzpicture}[scale=2]
\draw[black,thick,->] (0,0)-- (2,0) node [right] {$\omega$};
\draw[black,thick,->] (0,0)-- (0,2) node [right] {$t$};
\draw[black,dashed] (1,0)-- (1,2) node [right] {$\omega_0$};
\draw[black,dashed] (0,3/2)-- (2,3/2) node [right] {$t_d$};
\draw [red,thick,<-] (.8,3/2) -- (1.2,.3);
\draw [red,thick,<-] (.6,3/2) -- (1.,.3);
\draw [red,thick,<-] (.4,3/2) -- (.8,.3);
\draw [blue,fill] (.6,3/2) circle (.05);
\draw [blue,fill] (1.5-.5,.3) circle (.05);
\draw[black,thick,->] (4,0)-- (4+2,0) node [right] {$\omega$};
\draw[black,thick,->] (4,0)-- (4,2) node [right] {$t$};
\draw[black,dashed] (1+4,0)-- (1+4,2) node [right] {$\omega_0$};
\draw[black,dashed] (4+0,3/2)-- (4+2,3/2) node [right] {$t_d$};
\draw [red,thick] (5,3/2) --(5,.5);
\draw [red,thick,<-] (5-.2,3/2) --(5-.2,.5);
\draw [red,thick,<-] (5+.2,3/2) --(5+.2,.5);
\draw [blue,fill] (5,3/2) circle (.05);
\draw [blue,fill] (5,.5) circle (.05);
\end{tikzpicture}\caption{$K$ defines a Hamiltonian vector field on the time-frequency plane  shown as red arrows. $t_d$ is the detection time and $\omega_0$ 
the frequency of the source at emission time.   The boundary conditions at $\theta=\pm\pi$ select the orbit that starts on the vertical line $\omega_0$ and terminates 
on the horizontal  line $t_d$.  This is the orbit that connects the blue dots.  
The figure on the right shows the situation in the stationary case: $\omega$ is conserved as the vector field in phase space is vertical.}
\label{f:orbits} 
\end{figure}

The boundary conditions for the Hamiltonian system $(t,\omega)$  are  non-standard: Final boundary conditions are imposed on $t_\pm(\theta)$ while initial 
boundary conditions are imposed on $\omega_\pm(\theta)$.  
The mixture of initial and final boundary conditions is unusual from the perspective of mechanics and ODE in general\footnote{Mixed boundary conditions also show up in the semi-classical limit of quantum mechanics \cite{ls}}.  It is illustrated pictorially in Fig. \ref{f:orbits}.
 
Finally, we turn to the boundary conditions for the phase in the 
interferometer\footnote{In the case of a ring laser gyro, the boundary conditions are replaced by resonance conditions on the phase $\phi_\pm$.}.
The phase { $\phi_\pm$} of the $\pm$ waves are set at their emission times by the phase of the source. The boundary values for $\phi_\pm$ are then the initial values at the time of emission and are given by
	\be\label{e:initial}
	\phi_\pm(\mp\pi)
	= 
 {\Phi(t_{e\pm})}
	\ee
It follows {by Eq.~(\ref{e:phi})} that  the phase at the detector is then given by:
\begin{align}
\phi_\pm (\pm\pi)= \Phi(t_\pm(\mp\pi) )
-\int_{-\pi}^{\pi}{\cal L}(  t_\pm(\theta),\omega_\pm(\theta),\theta)
d\theta \label{e:phi1}
\end{align}
The phase difference of the $\pm$ waves at the detector is given by
\begin{align}
\Delta\phi &=\delta\Phi -\int_{-\pi}^{\pi}\left(\delta {\cal L}\right) d\theta  \label{e:Dphi} \end{align}
where
\be
\delta\Phi=\Phi(t_{e+} )-\Phi(t_{e-} )=-\omega_0' \int_{t_{e-}}^{t_{e+}}\frac{ds}{\gamma_\pi(s)}
\ee
and 
\be\label{yug}
\delta{\cal L}= {\cal L}(  t_+,\theta,\omega_+)-{\cal L}(  t_-,\theta,\omega_-)
\ee

In the framework of geometric optics $\phi_\pm$ are large quantities and  $\Delta\phi$, being the difference of two large quantities, should be handled with care.
As we see, when the fiber motion is adiabatic and $\beta\ll 1$,  the first term {in Eq.~(\ref{e:Dphi})}  is an integral over a short time interval  and the second has an almost 
self-canceling  integrand. 

\section{Stationary interferometers}\label{s:sta}  
We say that an interferometer is stationary {in the comoving coordinates $(t,\theta)$}
if
\footnote{
{A more general notion of stationarity may be defined by demanding existence of time like {Killing} vector field 
{on the world tube}.
 Since our analysis is tied to the comoving coordinates we 
only consider translation under $(\partial/\partial_t)_\theta$.
For a formulation in terms of arbitrary coordinates see appendix \ref{a:stationary}}}:
\be \label{68t6}
\omega_0=const, \quad {\partial_tK_\pm}=0
\ee
This setting corresponds to {having only} the term of order $O(\e^0)$ in the adiabatic expansion, i.e. to $\Delta^{(0)}$.

The condition $\omega_0=const$ expresses the stationarity of the light source.  The fiber is stationary in the $(t,\theta)$ coordinates if
$n,|\mbf{e}|,\gamma,\gamma_\perp$ and $\scp{e}{v}$ are time independent. This gives 
$\partial_tK=0$.  This holds for  Sagnac interferometer (of arbitrary shape) 
rotating like a rigid body with constant angular velocity and also for  {non-stretching} treadmill fiber interferometers\footnote{For planar fibers one may show that
actually these are the only two possibilities of stationary motions.} 
moving at constant speed,  such as the one shown in Fig. \ref{f:w}.
Stationarity is a strong condition and few examples satisfy it exactly. However, it is often  satisfied approximately. 
In such cases the stationarity assumption is useful as a basis for the adiabatic
approximation described in the next section.

\begin{figure}
\begin{centering}
\begin{tikzpicture}[scale=1.5]
\draw[ultra thick,red] (0,-.5)-- (3.3,-.5);
\draw[ultra thick,red] (0,.5)-- (3,1);
\draw[ultra thick,red]  (0,.5) arc (100:270:.5);
\draw[ultra thick,red]  (3,1) arc (105:-90:.77);
\draw [->,blue,thick] (3.2,-.3) arc (-90:90:16 pt);

\draw [->,blue,ultra thick] (1.3,-1.2)--(1.3,-.85);
\draw [<-,blue,ultra thick] (1.7,-1.2)--(1.7,-.85);
\draw[ultra thick,red,->] (1.8,-.5)-- (2.5,-.5);
\draw[ultra thick,red,->] (1.2,-.5)-- (.5,-.5);
\draw[fill] (1.5,-.5) circle [radius=.2];
\end{tikzpicture} 
\par\end{centering}

\caption{ The optical fiber moves {like a treadmill}
in the direction marked by the blue arrow. The black dot represents the co-moving beam-splitter
and the two vertical arrows the light source and detector. The counter-propagating
light-beams share a common path and are marked by red arrows. { If the fiber moves at constant speed without stretching, $\partial_tK_\pm=0$, and the interferometer is stationary in the sense of Eq.~(\ref{68t6})}. }
\label{f:w} 
\end{figure}
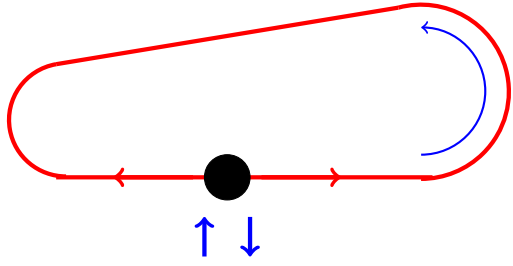

$\partial_tK_\pm=0$, implies by Eq.~(\ref{e:ham}), that $\omega_\pm(\theta) $ is constant on each trajectory. The boundary condition on 
$\omega_\pm$ then say that 
\be
\omega_+=\omega_-=\frac{\omega_0'}{\gamma_\pi}=\omega_0
\ee

  Since the source has constant frequency in the lab $\omega_\pm$ are also constant.
It  also  follows from $\partial_tK_\pm=0$ that ${\cal L}$  is time independent.  Consequently, {(see Eq.~(\ref{yug}))} 
$\delta{\cal L}$ vanishes identically and we are left with 
 \begin{align}
\Delta\phi &= -\omega_0' \int_{t_{e-}}^{t_{e+}}\frac{ds}{\gamma_\pi(s)}\nonumber \\
&=\frac{\omega_0' }{\gamma_\pi}\left({t_{e-}}-{t_{e+}}\right)\nonumber
\\
&=-\omega_0 \left(t_{e+}-t_{e-}\right)\nonumber \\
&=-\omega_0 \Delta t_e
\end{align}
\color{black}
where {$\Delta t_e$ is the difference in times of emmision and}
we made use of the fact that the source is in stationary motion. 
{
The elapsed  times between emission and detection is given by :
\be\label{e:phasegain}
(t_d-t_e)_\pm=
t_\pm(\pm\pi)-t_\pm(\mp\pi)= \int^{\pi}_{-\pi}\partial_{\omega} K_\pm  d\theta 
\ee
Since $K_\pm$ does not depend on $t$ or $\phi$ and since $\omega$ is a constant, the integrand is a {known} function of $\theta$.
}
Using the fact that the detection times for the $(\pm)$ {waves} are the same, $t_+(\pi)=t_d=t_-(-\pi)$,
the difference in the emission times  is given by
\begin{align}\label{e:delay}
\Delta t_e &=
t_+(-\pi)-t_-(\pi )\nonumber \\ &=
-\int_{-\pi}^{\pi} \left(\partial_{\omega}K_+ -\partial_{\omega} K_-\right) d\theta
\nonumber\\
&=
-2\int_{-\pi}^{\pi} \partial_{\omega}K_2\, d\theta
\nonumber\\
&=
-2\int_{-\pi}^{\pi}
  {\gamma}^2 {\bold v\cdot \bold{e}} \, d\theta
=-2\oint {\gamma}^2 {\bold v\cdot d\bold{\ell}} 
\end{align}
This completes the proof of Eq.~(\ref{e:stat}).

Geometric optics is concerned with the regime $\omega_0 \tau \gg 1$. 
Since 
\be\label{e:opt}
(\Delta\phi)^{(0)}=O\left( \beta \gamma^2\,(\omega_0\tau)\right) 
\ee
 the phase shift  depends sensitively on the velocity via $ \beta \gamma^2$.  
This is both a bug and a feature. It is a feature in the sense that phase shifts of $O(1)$  allow to measure the velocity with great accuracy. 
It can be bug because small fluctuations of the velocity can make the interference pattern unstable.   Ring interferometers become increasingly sensitive and eventually unstable at  relativistic velocities.  The optimal regime for the ring interferometers is when the velocity is  adjusted to the frequency $\omega_0$  so that the phase 
shift is $O(1)$.


\section{The adiabatic correction}\label{s:ad}

{
We say that the interefometer is adiabatic if the Hamiltonian $K_\pm$ and the source frequency $\omega_0'/\gamma_\pi$ change little during a cycle time $\tau\sim \ell/c$ 
of a light pulse through the fiber
\footnote{More systematically, the adiabatic limit is represented by writing $K_\pm$ and $\gamma_\pi$ as functions of scaled dimensionless time $s=\e t/\tau$ rather than of $t$}
\be
\big(\tau\partial_t \log(K_\pm) ,\ \tau\partial_t \log(\gamma_\pi)\big)\sim\e\ll1
\ee

We shall now consider the leading adiabatic corrections $\Delta^{(1)}$ which we will find to be  of order   
$O\left(\e\,(\omega_0\tau)\right)=O\left(\e\beta^0\,(\omega_0\tau)\right)$.
Comparison with Eq.~(\ref{e:opt}) shows that $\Delta^{(1)}$ may be as large as $\Delta^{(0)}$ even for $\e\ll1$ provided that {$\beta\ll 1$}. 
For this reason and for simplicity we shall assume in the following $\beta\ll1$.
In particular it allows neglecting $K_2=O(\beta\omega\tau)$ in comparison with 
$K_1=O(\omega\tau)$ { in Eq.~(\ref{e:H})}.


The adiabatic expansion of $(t,\omega)$ can be organized as:  
\be
t_\pm(\theta)=t_d+t^{(0)}_\pm(\theta)+t^{(1)}_\pm(\theta)+\dots,\quad
\omega_\pm(\theta)=\omega_0+\omega^{(1)}_\pm(\theta)+\dots
\ee
where
\be
t^{(0)}_\pm=O(\tau), \quad t^{(1)}_\pm=O(\e \tau), \quad \omega^{(1)}_\pm=O(\e \omega_0)
\ee
From  Eq.~(\ref{e:Dphi}) 
\be\label{e:dphi1}
 (\Delta\phi)^{(1)}= (\delta\Phi)^{(1)}-\int_{-\pi}^{\pi}\left(\delta {\cal L}\right)^{(1)}
d\theta
 \ee
{
We have already seen in the previous section (see Eq.(\ref{e:delay})) that:
\be\label{e:0t}
\Delta t_e^{(0)}=t^{(0)}_+(-\pi)-t^{(0)}_-(\pi)=O(\beta)
\ee
It follows that at leading order in $\beta$ we may neglect all powers of $\Delta t_e^{(0)}$ and write
\begin{align}\label{e:dphi1}
(\delta\Phi)^{(1)}&\approx -\omega_0 \Delta t_e^{(1)}=
  -\omega_0 \big(t_+^{(1)}(-\pi)-t_-^{(1)}(\pi)\big)
\end{align}
}
{Since $\tau\partial_t{\cal L} /{\cal L}=O(\e)$ while $\omega_0\partial_\omega{\cal L}/{\cal L} =O(1)$ we have}
\be\label{e:dL1}
\delta{\cal L}^{(1)}\approx (\partial_t {\cal L})\, \delta t^{(0)}+(\partial_\omega {\cal L})\, \delta \omega^{(1)}
\ee
{where} $\partial {\cal L}=(\partial {\cal L})(t_d,\theta,\omega_0)$ are evaluated at the detection time and
\be
\delta t^{(0)}(\theta)=t^{(0)}_+(\theta)-t^{(0)}_-(\theta),\quad
\delta \omega^{(1)}(\theta)=\omega^{(1)}_+(\theta)-\omega^{(1)}_-(\theta)
\ee
To compute $(\Delta\phi)^{(1)}$ we therefore need $t_\pm^{(1)}$ at the emission point and $( \delta t^{(0)}, \delta \omega^{(1)})$ along the fiber. 
Interestingly, as we shall see below, we shall not need $\omega^{(1)}$.  

$t_\pm^{(0)}$ has two pieces: The dominant piece that is $O(\tau)$ which comes from $K_1$ and a piece of order $O(\tau \beta)$ that comes from $K_2$.  
It is the smaller piece that gives the Sagnac effect at order $\e^0$. For evaluating Eq.~(\ref{e:dL1}) {to leading order in $\beta$}
we only need the dominant part of $t_\pm^{(0)}$   which 
is the solution of 
\begin{align}\label{e:series0}
\pm{dt^{(0)}\over d\theta}&\approx \partial_\omega K_ 1(t_d,\theta,\omega_0)\end{align}
 Hence
\be
t_\pm^{(0)}(\theta)\approx \mp \int_\theta^{\pm\pi}\partial_\omega  K_1 (t_d,\theta',\omega_0)d\theta'
\ee
 This gives
\be\label{e:dt0}
\delta t^{(0)}(\theta)\approx \int^\pi_{-\pi}sgn(\theta-\theta')\partial_\omega  K_1(t_d,\theta',\omega_0)d\theta'
\ee
We need $t_\pm^{(0)}$ to find $t_\pm^{(1)}$ which is  the solution of 
\begin{align}\label{e:series}
\pm{dt^{(1)}\over d\theta}&\approx(t^{(0)}\partial_t+\omega^{(1)}\partial_\omega) \partial_\omega K_1(t_d,\theta,\omega_0)
\end{align}
Integrating Eq.~(\ref{e:series}) we find
\begin{align}\label{e:w1}
(\delta\Phi)^{(1)}&=-\omega_0\left(t^{(1)}_+(-\pi)-t^{(1)}_-(\pi)\right)\nonumber \\ &=\omega_0\int_{-\pi}^\pi d\theta   
\left(
\delta t^{(0)}(\theta)\partial_t 
 +\delta\omega^{(1)}(\theta)\partial_\omega 
\right)
\partial_\omega K_1(t_d,\theta,\omega_0)
\end{align}
  $(\Delta\phi)^{(1)}$ has four terms: Two  that come 
from the integral in Eq.~(\ref{e:w1})  and {two } from integrating Eq.~(\ref{e:dL1}). Summing these four terms give
\be
(\Delta\phi)^{(1)}=
\int \, d\theta \Big(\delta\omega^{(1)}(\theta)\big( 
\omega\partial_{\omega}^2 K-\partial_\omega
{\cal L}\big)+
\delta t^{(0)}(\theta)\big( 
\omega\partial_{\omega}\partial_{t} K-\partial_t
{\cal L}\big)\Big)
\ee
The first bracket vanishes   since
\be
\omega\partial_{\omega}^2 K-\partial_\omega
{\cal L}=\omega\partial_{\omega}^2 K-\partial_\omega(\omega\partial_\omega K)+\partial_\omega K=0
\ee
The second bracket is
\be\label{e:ptK}
\omega\partial_{\omega}\partial_{t} K-\partial_t
{\cal L}=\omega\partial_{\omega}\partial_{t} K-\partial_t
(\omega\partial_\omega K-K)=\partial_tK
\ee
The term proportional to $\delta\omega^{(1)}$ drops and the term proportional to $\delta t^{(0)}$ survives.  
  Substituting Eq.~(\ref{e:dt0}) and using Eq.~(\ref{e:ptK})  gives:
\begin{align}\label{e:dphi1f}
(\Delta\phi)^{(1)}&=\int_{-\pi}^\pi \int_{-\pi}^\pi  d\theta \,d\theta' \sgn(\theta-\theta')\partial_t K_1(\theta)\partial_\omega K_1(\theta')\nonumber \\
&\approx\frac\omega {c^2}\int_{-\pi}^\pi \int_{-\pi}^\pi  d\theta \,d\theta' \sgn(\theta-\theta')\partial_t\big(n|\mbf{e}|\big)(\theta)\partial_\omega \big(\omega n |\mbf{e}|\big)(\theta')
\\
&\approx\frac\omega {c^2}\int_{-\pi}^\pi \int_{-\pi}^\pi  |\mbf{e}(\theta)|d\theta \, |\mbf{e}(\theta')|\,d\theta' \sgn(\theta-\theta')\big(\partial_t n(\theta)+n(\theta) \sigma(\theta)\big)\ \partial_\omega \big(\omega n(\theta') \big)\nonumber
\\
&=\frac\omega {c^2}\int_{0}^L \int_{0}^L  d\ell \,d\ell' \sgn(\ell-\ell'))\big(\partial_t n(\ell)+
\partial_t\ell\, \partial_\ell n(\ell)+n(\ell) \sigma(\ell)\big) \partial_\omega \big(\omega n(\ell') \big)\nonumber
\end{align}
For typographical simplicity we  have omitted throughout the arguments $t$ and $\omega$ throughout.
The second line is the $\beta\ll 1$  approximation 
$K_1\approx \frac \omega c n|\mbf{e}|$.    
In the third line we used the non-relativistic approximation for the stretch rate $\sigma$, Eq.~(\ref{e:sig2}).  
{In the last line we changed variables from $\theta$ to $\ell$ so that $n(\theta,t)$ is replaced by $n(\ell,t)$.} 
{$L$ is the length of the fiber at the time of detection $t$ and $\partial_t\ell= \int_0^\ell \sigma(\ell')d\ell'$.}  
This proves Eq.~(\ref{e:2nd}).

\section{Applications}
\subsection{Thermal non-reciprocity}\label{s:shupe}
Consider a {static} homogeneous fiber with a temperature profile $T(\theta,t)$ whose variation $\delta T$ is small.  The index of refraction is  assumed to be a function of the temperature, $ n(T,\omega)$,  and similarly the length $d\ell$ is a function of $T$, i.e.  $d\ell= |\mbf{e}(T)| d\theta$. The coefficient of thermal expansion is then $\alpha= \partial_T\log |\mbf{e}(T)|$. We are interested in computing $\Delta \phi^{(1)}$ to first order in $\delta T$. This problem has been considered in \cite{shupe}.

 {Evidently}
 \be
\partial_t (n|\mbf{e}|) =\partial_T (n|\mbf{e}|)\, \partial_t(\delta T)
 \ee
This implies that the first brackets in Eq.~(\ref{e:dphi1f}) is $O(\delta T)$ and we may 
{ignore the $\delta T$ dependence of the other factors. For a homogeneous fiber} the second brackets {is} a constant which can be pulled out of the integral. 
\begin{align}
(\Delta\phi)^{(1)}
&\approx\frac\omega {c^2}\int_{-\pi}^\pi \int_{-\pi}^\pi  d\theta \,d\theta' \sgn(\theta-\theta')\partial_t\big(n|\mbf{e}|\big)(\theta)\partial_\omega \big(\omega n |\mbf{e}|\big)(\theta')\nonumber
\\
&\approx\frac\omega {c^2}
\partial_T\big(n|\mbf{e}|\big)\partial_\omega \big(\omega n |\mbf{e}|\big)\int_{-\pi}^\pi \int_{-\pi}^\pi  d\theta \,d\theta' \sgn(\theta-\theta')\partial_t(\delta T)\nonumber\\
&\approx 2\frac\omega {c^2}
{|\mbf{e}|\big(\partial_Tn+n \alpha\big)}
\partial_\omega \big(\omega n |\mbf{e}|\big)\int_{-\pi}^\pi  d\theta \,\theta\, \partial_t(\delta T)
\end{align}
Changing variables from $\theta$ to {length $\ell=|\mbf{e}|(\theta+\pi)$} we can write 
\begin{align}
(\Delta\phi)^{(1)}
&\approx \frac\omega {c^2}
\big(\partial_Tn+n \alpha\big)\partial_\omega \big(\omega n\big)
\int_{0}^L  d\ell \,(2\ell-L) \big(\partial_t(\delta T)+\partial_t\ell \, \partial_\ell (\delta T) \big)
\end{align}
However, {for static fibers $\partial_t\ell$ vanishes apart from a  $\sim\alpha \delta T$  correction
which is negligible to leading order in $\delta T$.}
Hence to first order in $\delta T$
 \begin{align}
(\Delta\phi)^{(1)}
&\approx \frac\omega {c^2}
\big(\partial_Tn+n \alpha)\big)\partial_\omega \big(\omega n\big)\int_{0}^L  d\ell \,(2\ell-L) \big(\partial_t(\delta T)\big)
\end{align}
 {This is a slight generalization of Shupe formula \cite{shupe} in that it accounts also for the case that $n$ is dispersive. }

   Note that if the spatial distribution of $\partial_t(\delta T)$  {has short range correlations (idealized by white noise)}
 then 
   \be
   \average{\left((\Delta\phi)^{(1)}\right)^2}=O(L^3)
   \ee
which, {for large $L$}, dominates the Sagnac  $O(L)$ term which is.
 
\subsection{Fizeau}\label{s:fizeau}

The Fizeau experiment is shown schematically in  Fig.~\ref{f:fizeau}.  The interferometer is at rest in the lab, but one arm of the interferometer has a flowing liquid, say, water.  
From the perspective of the lab the setting is stationary, as the velocity at each point in the lab frame is time independent. However, from the point of view of  the co-moving coordinate $\theta$, the interferometer is not stationary: Elements of the fluid accelerate and decelerate  as
 they enter and leave the pipe. 
{Fig. \ref{f:2} compares the lab $(x,t)$ and $(\theta,t)$ coordinates.}  

In the lab coordinates the velocity is approximately  given by
\be\label{e:delta}
v(x)= V\chi_{(x_{in},x_{out})}(x) 
\ee
{where $\chi_{(x_{in},x_{out})}$ is the characteristic function of the interval. The stretching is then}
\be\label{swq}
\sigma=\partial_t \log |\mbf{e}|=\frac{\partial_\theta v}{|\mbf{e}|}=\partial_x v=
V(\delta(x-x_{in})-\delta(x-x_{out}))
\ee
Substituting {Eq.~(\ref{e:delta})} in  Eq.~(\ref{e:stat}) 
(and using {$|\mbf{e}| d\theta=dx$}) gives for the Sagnac term:
\be
(\Delta\phi)^{(0)}=-2\omega VL
\ee
Substituting {Eq.~(\ref{swq})} in   Eq.~(\ref{e:2nd})  gives the correction due to stretching:
\be
(\Delta\phi)^{(1)}=n\omega \partial_\omega(\omega n)\int \sigma(x)\sgn(x'-x)dxdx'
=n\omega \partial_\omega(\omega n)2VL.
\ee
Summing we obtain von Laue result,  Eq.~(\ref{e:f}).

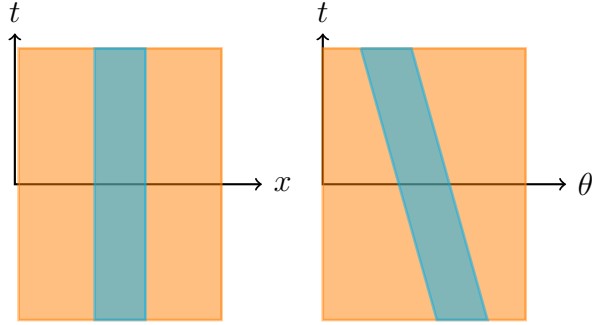
\begin{figure}[h!]
\centering{} 
\begin{tikzpicture}[scale=1,line width=1pt,black]
 \draw [<->,thick](3.2,0)--(-.05,0)--(-.05,2) node [above] {$t$};
 \node [right] at (3.2,0) {$x$};
  \draw [fill, orange,opacity=.5] (0,-1.8) rectangle (8/3,1.8);
   \draw [fill, cyan,opacity=.5] (3/3,-1.8) rectangle (5/3,1.8);

   \begin{scope}[shift={(4,0)}]
    \draw [<->,thick](3.2,0)--(0,0)--(0,2) node [above] {$t$};
 \node [right] at (3.2,0) {$\theta$};
 \draw [fill, orange,opacity=.5] (0,-1.8) rectangle (2/3+6/3,1.8);
 \draw[fill, cyan,opacity=.5] (1/2, 1.8)--(7/6,1.8)--(13/6,-1.8)--(3/2, -1.8)--(1/2, 1.8);
\end{scope}
   
  \end{tikzpicture} 
\caption{The velocity in in the right arm of Fizeau. Shown on the left is the velocity of the fluid as a function of the lab coordinates $(t,x)$.   The orange strip represents zero velocity and the cyan velocity $V$. In the lab frame the velocity is independent of $t$  and the setting is stationary.  The diagram on the right illustrates  the velocity field in the $(t,\theta)$ coordinates. The velocity is non zero in  the (cyan) parallelogram. We assume that a fluid particle starting at the entrance boundary, discontinuously changes  its velocity from zero to $V$, and then stops abruptly at the exit boundary.  The velocity is time dependent, reflecting non-stationarity  in the $(t,\theta)$ coordinates.   }
\label{f:2} 
\end{figure}

\color{black}

\color{black}

\section*{Acknowledgment}

The research is supported by ISF. We are especially grateful to Amos Ori for getting us interested in this problem and for many instructive and insightful discussions. 
We thank B. Shapiro for telling us about \cite{menegozzi,shiozawa}, and M. Tur for a useful conversation. Fig. \ref{f:3}, \ref{f:fizeau} and  \ref{f:w}  are taken from \cite{Ori}.
{JEA gratefully acknowledges support from the Simons Center for Geometry and Physics, Stony Brook University at which some of the research for this paper was performed.}

\appendix
\section{Moving fibers in arbitrary coordinates}\label{a:stationary}

\subsection{A world sheet in general coordinates }

A closed optical fiber deforming in time traces out a two dimensional surface
in space-time which, in analogy to string-theorist terminology, may be referred
as a `world sheet'.
One may describe it parametrically as $X=X^\mu(t,s)$.
$t$ is a time-like coordinate 
and $s$ is a space-like periodic coordinate (i.e. $\dot{X}^\mu=\partial_t X^\mu$ 
and ${X'}^\mu=\partial_s X^\mu$ are time-like and space-like respectively).
The refraction index of the fiber is $n(t,s,\omega)$.

To fully describe the fiber one must also be given its local 4-velocity.
This is given by a tangent vector field
\footnote{{The world sheet tangent vectors $v,e$ defined here should not be confused with the spatial vectors mentioned in the main text.}}
 $v$ on the world sheet.
We write it using two components $v=(v^0,v^1)$. 
The (unit) 4-vector corresponding to it is 
\be
v^i\partial_i X^\mu=v^0 \dot{X}^\mu+v^1 {X'}^\mu
\ee
The three functions $X^\mu(t,s),v^\mu(t,s),n(t,s,\omega)$
suffice to fully describe our system.

It is convenient to define another tangent vector $e=(e^0,e^1)$ by demanding it
to be orthogonal to $v$ and normalized. (Its orientation may be chosen by demanding $e^1>0$.)
It is possible to write $e$ explicitly in terms of $v$, the Levi-Civita tensor $\varepsilon$,
and the metric $g_{ij}=\eta_{\mu\nu}\partial_i X^\mu \partial_j X^\nu$.
The pair $(v,e)$ form a 2-dim vierbein on the world sheet.

\subsection{The Eikonal equation}
Within the eikonal approximation, a wave moving in the fiber is 
described by its phase $\phi(t,s)$.
The corresponding frequency and wave vector are than defined to be
\be
\omega=-\partial_t\phi,\;\;\;\;\; k=\partial_s\phi
\ee
The standard relation $k'=\pm n\omega'$ is satisfied however by the wave-vector and frequency
$\omega',k'$ in the local inertial frame $(v,e)$ moving with the fiber.
\be
\omega'=-v^i\partial_i\phi=v^0\omega-v^1 k,\;\;\;\;
k'=e^i\partial_i\phi=-e^0\omega+e^1 k
\ee
In the absence of chromatic dispersion one easily solves $k'=\pm n\omega'$ obtaining
 \begin{equation}\label{kw}
  k=\omega{e^0 \pm n v^0\over e^1 \pm n v^1 }
  \end{equation}
When considering dispersive media one must keep in mind that 
 the refraction index $n$ appearing here
should be evaluated at the local moving frame, i.e.
\be
n=n(\omega')=n(v^0\omega-v^1 k)
\ee
Thus in order to really find $k(\omega)$, may require solving a nontrivial equation.
This difficulty is avoided if one chooses $s$ to be a comoving coordinate
(such that each value of $s$ corresponds to a specific fixed material point)
since then $v^1\equiv 0$ and hence $n=n(v^0\omega)$ does not depend on $k$.
Also in nonrelativistic limit one typically has $v^\sigma\ll v^\tau$ 
(unless the coordinates $\tau,\sigma$ are chosen in a very unnatural way).
Expanding in powers of $\beta$ than allows a perturbative solution for $k(\omega)$.

The two solutions of Eq. (\ref{kw}) define two functions of $\omega,t,s$
which we shall denote by $\pm K_{\pm}(t,s,\omega)$.
(The extra sign in front of $K$ is needed for consistency with the rest of our conventions.
Under  non-extreme circumstances it corresponds to having both $K_+$ and $K_-$ positive.)

Having found the functions $K_\pm$, the Hamilton equations for the eikonal wave propagation
can be written in the standard form.
\be
\pm \frac{dt}{ds}=\partial_\omega K_\pm,\;\;\; 
\pm\frac{d\omega}{ds}=-\partial_t K_\pm
\ee
\be
\frac{d\phi}{ds}=\mp{\cal L}_\pm,\;\;\;
{\cal L}_\pm=\omega\partial_\omega K_\pm-K_\pm
\ee

\subsection{Consistency with Eq.~(\ref{e:H}) }
Choosing $s=\theta$ a comoving coordinate means that the matter particles move in the direction defined by $\partial_t$ and hence $v^1\equiv 0$.
If we further assume that $t$ is the lab time such that $X=(t,\mbf{x}(t,\theta))$
than we may also solve for $v^0,e^0,e^1$ obtaining 
\be 
v^0=\gamma,\; e^1=\frac{\gamma_\perp}{\gamma |\mbf{x}'|},\; 
e^0=\gamma\gamma_\perp {\dot{\mbf{x}}}_\|
\ee
This  reproduces $K_\pm $ of Eq.~(\ref{e:H}).

$$K_\pm={\gamma^2\over\gamma_\perp}n\omega |\mbf{x}'|\pm\gamma^2\omega (\dot{\mbf{x}}\cdot \mbf{x}')$$

\subsection{The notion of stationarity}
The fiber and the associated interferometer may be called stationary if there exists
a timelike vector field $\xi$ defined on the world sheet under which the system is symmetric.
This means that the $\xi$-Lie derivative of $v,n,g_{ij}$ vanish.
Note that this is equivalent to vanishing of the $\xi$-Lie derivative of $v,e,n$.
In section \ref{s:sta} we considered only the case of $\xi=v$ {(which made the condition $L_\xi v=0$ trivial)}.
The simplest nontrivial example of a stationary interferometer with $\xi\neq v$ is Fizeau.

\subsection{Fizeau yet again}\label{a:laue}

The Fizeau experiment is described most simply in terms of the lab coordinates $(t,x)$
where it is {explicitly time independent}.
Denoting the fluid velocity by $V$ (which may depend on $x$ but not on $t$)
we  have
\be 
v=(\gamma,\gamma V),\;\;\; e=(\gamma V,\gamma)
\ee
with
\be
K_\pm=\omega {n(\omega')\pm V\over 1\pm Vn(\omega')}
\ee
Noting that in the nonrelativistic limit
$$\omega'=\gamma(\omega-Vk)=\omega\mp V \omega n(\omega)+O(\beta^2)$$
We obtain
$$K_\pm=n_0\omega\pm (1-n_0^2)V\omega\mp V\omega^2 n_0\partial_\omega n_0+O(\beta^2)$$
where we denoted $n_0=n(\omega)$ in order to distinguish it from $n(\omega')$.

Since $K_\pm$ does not depend on time, it is easy to solve Hamiltons equations.
$$\pm d\omega/dx=-\partial_t K_\pm=0 \Rightarrow \omega_{\pm}=\omega_0$$
$$\pm dt_{\pm}/dx=\partial_\omega K_\pm \Rightarrow (t_f-t_i)_\pm=\int dx \partial_\omega K_\pm$$
$$\pm d\phi_\pm/dx =-{\cal L}_\pm \Rightarrow (\phi_f-\phi_i)_\pm=-\int dx{\cal L}_\pm$$
Adding both contribution we find the phase at arrival at the detector is 
$$(\phi_f)_\pm+\omega t_f=(\phi_f-{\phi_i})_\pm+\omega(t_f-{t_i})_\pm=
\int_{x_i}^{x_f} dx K_\pm.$$
This result was expected and could have been easily derived without use of our general formalism.
The phase difference at arrival is then
\begin{align}
\Delta\phi&=\int_{x_i}^{x_f} dx (K_+-K_-)\nonumber \\
&=
2(\omega(1-n_0^2)-\omega^2 n_0\partial_\omega n_0)\int V dx+O(\beta^2)\nonumber
 \\
&=
2\omega\big(1-n_0\partial_\omega(\omega n_0)\big)\int V dx+O(\beta^2)
\end{align}

\end{document}